\documentclass[letterpaper,english,twocolumn, aps,superscriptaddress]{revtex4-1}
\usepackage{color}
\usepackage{amsmath}
\usepackage{amssymb}
\usepackage{graphicx}
\usepackage{cleveref}

\makeatletter

\pdfpageheight\paperheight
\pdfpagewidth\paperwidth

\newcommand{\rev}[1]{\textcolor{black}{#1}}

\makeatother

\usepackage{babel}

\begin{document}
\title{Cloaking Transition of Droplets on Lubricated Brushes}

\author{Rodrique G. M. Badr}
\email{rbadr@uni-mainz.de}
\affiliation{Institut f\"ur Physik, Johannes Gutenberg-Universit\"at Mainz, Staudingerweg
	7, D-55099 Mainz, Germany}

\author{Lukas Hauer}
\affiliation{Max Plank Institut f\"ur Polymer Forschung Mainz, Ackermannweg 10, 55128 Mainz, Germany}

\author{Doris Vollmer}
\affiliation{Max Plank Institut f\"ur Polymer Forschung Mainz, Ackermannweg 10, 55128 Mainz, Germany}

\author{Friederike Schmid}
\email{friederike.schmid@uni-mainz.de}
\affiliation{Institut f\"ur Physik, Johannes Gutenberg-Universit\"at Mainz, Staudingerweg
	7, D-55099 Mainz, Germany}

\begin{abstract}

 We study the equilibrium properties and the wetting behavior of a simple liquid on a polymer brush, with and without presence of lubricant \rev{by multibody Dissipative Particle Dynamics simulations. The lubricant is modelled as a polymeric liquid consisting of short chains that are chemically identical to the brush polymers}. We investigate the behavior of the brush in terms of the grafting density and the amount of lubricant present. Regarding the wetting behavior, we study a sessile droplet on top of the brush. \rev{The droplet consists of non-bonded particles that form a dense phase}. Our model and choice of parameters result in the formation of a wetting ridge and in the cloaking of the droplet by the lubricant, \rev{i.e. the lubricant chains creep up onto the droplet and eventually cover its surface completely. Cloaking is a phenomenon that} is observed experimentally and is of integral importance to the dynamics of sliding droplets. We quantify the cloaking in terms of its thickness, which increases with the amount of lubricant present. The analysis reveals a well-defined transition point where the cloaking sets in. We propose a thermodynamic theory to explain this behavior. In addition we investigate the dependence of the contact angles on the size of the droplet and the possible effect of line tension. \rev{We quantify the variation of the contact angle with the curvature of the contact line on a lubricant free brush and find a negative value for the line tension. Finally we investigate} the effect of cloaking/lubrication on the contact angles and the wetting ridge. \rev{We find that lubrication and cloaking reduce the contact angles by a couple of degrees. The effect on the wetting ridge is a reduction in the extension of the brush chains near the three phase contact line, an effect that was also observed in experiments of droplets on crosslinked gels.}

\end{abstract}
\maketitle

\section{Introduction}

Droplets are omnipresent. Important applications include self cleaning \citep{blossey2003self,nakajima2000transparent,parkin2005self}, spray coating, anti-icing, anti-fouling \citep{epstein2012liquid},
anti-corrosion or more eﬃcient application of pesticides\citep{bergeron2000controlling,gaume2002function}. Understanding the interactions of droplets with surfaces is also of inherent interest for physicists due to the interplay of various forces and energies acting at different length and time scales. Here, we aim to understand the wetting of water droplets on dry and lubricated brushes. \rev{Applications of polymer brushes in regards to wetting include moisture harvesting using PNIPAAm brushes on cotton fabric \citep{yang2013temperature}, anti-fogging using stimuli-responsive brushes \citep{howarter2007self}, and the manufacture of materials with modifiable and switchable adhesive, dissipative, and 
wettability properties \citep{ritsema2022fundamentals}. }

For a long time, wetting research focused on modelling the static and dynamic properties of droplets on smooth and rigid surfaces. Modelling wetting dates back to Thomas Young \citep{young1805iii}. He showed that in thermodynamic equilibrium the contact angle of a droplet deposited on an ideally smooth, chemically homogeneous rigid surface is determined by the interplay of the interfacial tensions.
\begin{equation}
\label{eqn:young}
 \cos\theta = \frac{\gamma_{s} - \gamma_{sw}}{\gamma_{w}}
\end{equation}
$\gamma_{s}$, $\gamma_{w}$ and $\gamma_{sw}$, are the solid-vapour, droplet-vapour and solid-droplet interfacial tension. While this equation holds for micrometer sized droplets and larger ones, for nanometer sized droplets line tension changes the contact angles, too. 
In contrast to Thomas Young’s equation, in real life a droplet never has a unique well-defined contact angle. Surface roughness, chemical inhomogeneities, adaptation of the surface due to the presence of a drop \citep{butt2018adaptive} or droplet induced charging of the surface \citep{stetten2019slide}
cause pinning of the three phase contact line. The measured apparent contact angle can take any value between the apparent receding and apparent advancing contact angle. Contrary to experiments, in simulations the realization of the ideal condition is possible. Modelling idealized conditions has the advantage that the influence of different factors on wetting phenomena can be well separated. In particular if small length or time scales are involved this might be hard or even impossible to probe with currently available experimental techniques. Here, it should be kept in mind that the wetting properties are determined by the properties of the droplet and the surface close to the three-phase contact line.

Particularly challenging is the understanding of the wetting properties of sessile droplets on dry or lubricant infiltrated brushes. A brush is composed of chains grafted by one end to a surface. The resulting layer is elastic and of very small thickness (order of nm). Therefore, the length scales are too small and curvatures are too large to experimentally investigate the shape of the area close to the three-phase contact line. However, to a certain extent, information gained from droplets on rigid surfaces, gels, or lubricant infused surfaces can be used to understand the wetting properties of droplets on dry and lubricated brushes \citep{smith2013droplet,wong2011bioinspired,quere2005non,lafuma2011slippery}. For example it has been shown that sessile droplets on gels and lubricant infused surfaces are surrounded by an annular wetting ridge. The vertical component of the interfacial tension at the three phase contact line exerts a force on the surface, pulling the gel or lubricant up. Depending on the elastic modulus of the gel or the thickness of the lubricant layer, now the contact angles can be modelled by a Neumann triangle \citep{marchand2012contact}. For lubricant impregnated surfaces the droplet can be covered by a thin cloak. Whether it is possible to form a cloak or not depends on the balance of the interfacial tensions. A cloak forms for a positive spreading coefficient $S$, 
$S= \gamma_{w} – \gamma_{o} - \gamma_{ow}$,
where $\gamma_{o}$ and $\gamma_{ow}$ is the interfacial tension of lubricant-vapour and droplet-lubricant, respectively. Interference measurements allow an estimate of the thickness of the cloak but are insufficiently sensitive to quantify the variation of the thickness over the surface\cite{kreder2018film}.
The reason is that the curvature of the droplet complicates data analysis. So far it is also unclear how the thickness of the cloak depends on the amount of lubricant available. 

\begin{figure}[!htb]
	\begin{centering}
		\includegraphics[width=8cm]{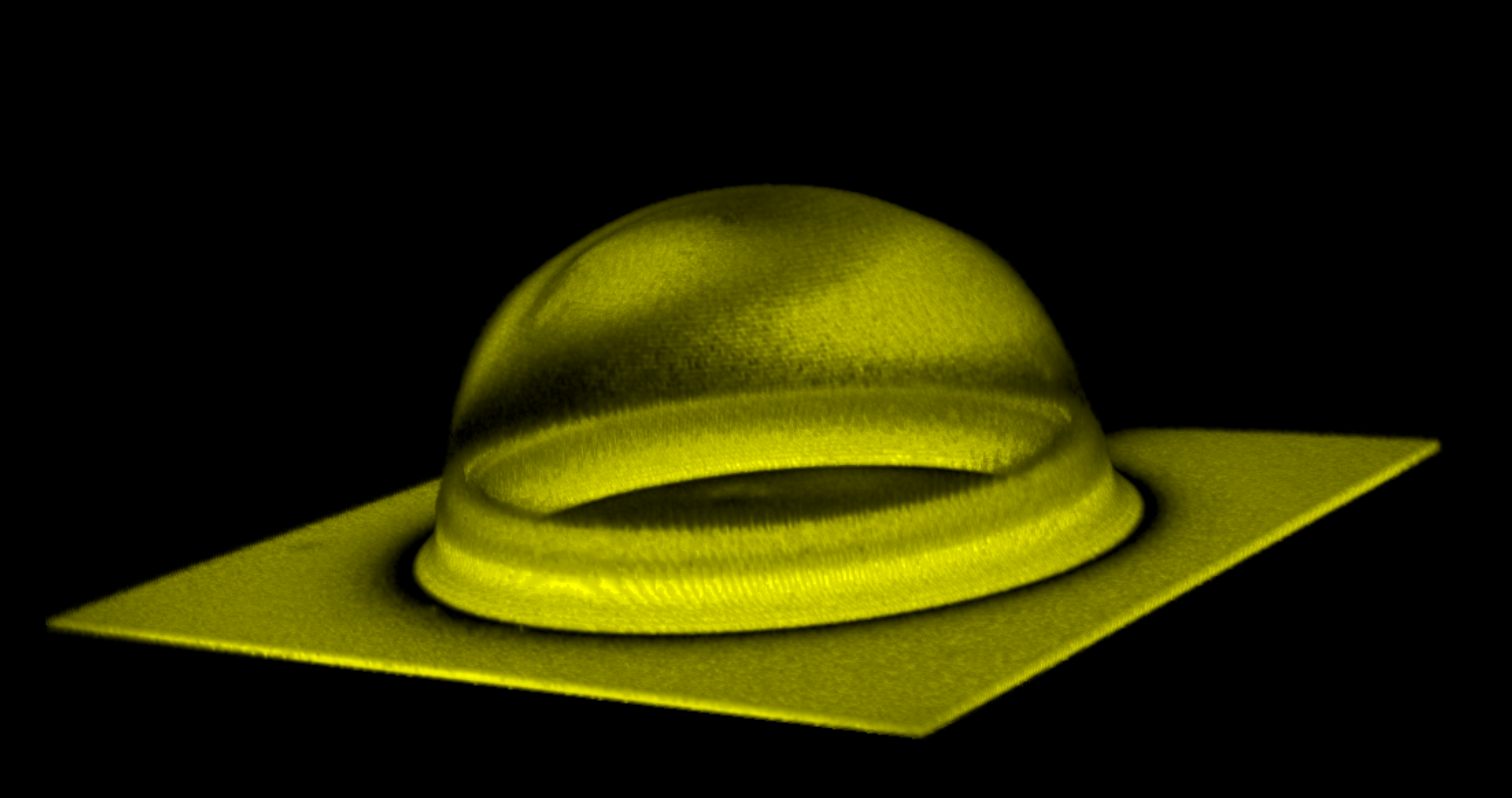}
        \includegraphics[width=8cm]{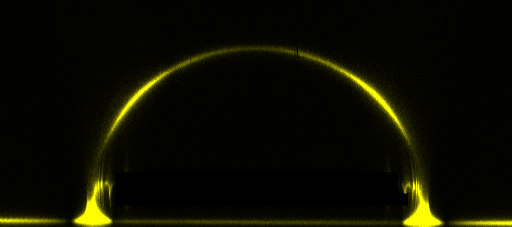}
	\par\end{centering}
	\caption{Top: Experimental 3D view of the droplet on a PDMS brush infused with silicone oil, obtained through confocal microscopy. The image shows a clear wetting ridge at the three phase contact line, in addition to a cloaking layer on the surface of the droplet. Bottom: Cross sectional side view of the same droplet. A wetting ridge is clearly visible, in addition to a cloak indicated by the fine fluorescence signal all over the droplet.}
	\label{fig:3DviewExp}
\end{figure}

In the present paper, we aim to tackle these questions for sessile droplets on dry and lubricated brushes using coarse-grained molecular dynamic simulations. We use a simple rather generic model, however, with interactions parameters adjusted to polydimethylsiloxane (PDMS).  Brushes made of PDMS recently have attracted a lot of attention because droplets experience particularly low contact line pinning \citep{krumpfer2011rediscovering}. Reported values\cite{wooh2016silicone, wang2016covalently,teisala2020grafting} for the contact angle hysteresis of water on PDMS vary between $1-11^\circ$. The differences may be caused by varying amounts of non-crosslinked chains in the brush. The reason for this low contact angle hysteresis is that PDMS chains are very ﬂexible, having a persistence length of a few monomers. The large bond angle of approximately $150^\circ$ between Si-O-Si atoms induces a very low torsional barriers \citep{weinhold2011nature}. The hydrophobic methyl side groups lead to the low interfacial tension of 0.02 N/m. They also shield the inorganic Si-O backbone \citep{colas2005silicones}. As opposed to PDMS gels where the chains are crosslinked resulting in an elastic solid, PDMS brushes keep their high flexibility. Figure \ref{fig:3DviewExp} shows an example of an experimental view of a droplet on a PDMS brush covered by a layer of silicone oil as a lubricant, obtained with confocal microscopy, \rev{where the yellow fluorescence signal indicates the presence of lubricant}. Details on the experimental setup are given in the Appendix A. The confocal image shows two very important features. Near the three phase contact line there is an accumulation of lubricant, forming a wetting ridge. Especially clear in the cross sectional view is a visible fluorescent signal all over the droplet, showing that the droplet is indeed cloaked by the lubricant.

While a molecular understanding of wetting properties of sessile droplets on brushes cannot be achieved experimentally, molecular dynamics studies can give detailed insight at the molecular level. Previous molecular dynamics simulations of polymeric droplets on polymer brushes by Léonforte and Müller have provided insight into the dependence of the contact angle and height of the wetting ridge on the affinity between the droplet and the brush, in addition to the effect of line tension \citep{leonforte2011statics}. Later work by Mensink, \rev{de Beer}, and \rev{Snoeijer} investigated the transition between mixing, total wetting, and partial wetting, of polymeric droplets on polymer brushes \citep{mensink2019wetting}, and studied the effects entropic contributions have on the transitions \citep{mensink2021role}. As for droplet on lubricated surfaces. Molecular dynamics simulations on lubricant-infused patterned surfaces have reproduced the phenomenon of the cloaking of the droplet by the lubricant \citep{guo2019droplet,guo2020dynamic}. Here, we use such simulations to study the shape of the \rev{wetting} ridge, the onset and thickness of the cloak, and the influence of the amount of lubricant on the apparent contact angle.

\section{Model and Methods}
\label{sec:sim}


\subsection{Model}

We consider a coarse-grained model system containing a polymer brush, free lubricant chains, and a droplet made of liquid particles, in coexistence with a vapor phase. Polymers are chains of beads connected by springs with the spring potential $U_{bond}=k(r_{ij}-r_0)^2$. Liquid molecules are modelled as single isolated beads. To model the non-bonded potential and the coupling to a heat bath at given temperature $T$, we use the Many-body Dissipative Particle Dynamics (MDPD) coarse grained model and thermostat\cite{pagonabarraga2002mdpd, trofimov2002mdpd, warren2003vapor}. The DPD thermostat has the advantage that the dissipative and random forces are pairwise interactions and momentum conserving, while the multi-body force element allows for the coexistence of two phases. The forces take the following form:

\begin{align}
	\begin{split}
		&F_{ij}=F_{ij}^C+F_{ij}^D+F_{ij}^R\\
		&F_{ij}^C=A_{ij}w^C(r_{ij})\hat{r}_{ij}+B_{ij}(\bar{\rho}_i+\bar{\rho}_j)\tilde{w}^C(r_{ij})\hat{r}_{ij}\\
		&F_{ij}^D=-\zeta w^C(r_{ij})^2(\hat{r}_{ij}.\vec{v}_{ij})\hat{r}_{ij}\\
		&F_{ij}^R=\sqrt{2\zeta k_B T}w^C(r_{ij})\theta_{ij}\hat{r}_{ij}\\
		&w^C(r_{ij})=\begin{cases}
			\bigg(1-\frac{r_{ij}}{r_c}\bigg)~~&r_{ij}\leq r_c\\
			0~~&r_{ij}>r_c
		\end{cases} \\
		&\tilde{w}^C(r_{ij})=\begin{cases}
			\bigg(1-\frac{r_{ij}}{r_d}\bigg)~~&r_{ij}\leq r_d\\
			0~~&r_{ij}>r_d
		\end{cases}\\
		&\bar{\rho}_i=\sum_{j\neq i}\frac{15}{2\pi r_d^3}\tilde{w}^C(r_{ij})^2
	\end{split}
\end{align}

\noindent In the above $F_{ij}^C$ is the conservative force contribution where $A_{ij}<0$ is the strength of the attractive part, and $B_{ij}>0$ is the strength of the density dependent repulsion. $B_{ij}$ must have the same value for all pairs of particles for the forces to be conservative as shown by the no-go theorem of MDPD \citep{warren2013no}. We also have $\vec{r}_{ij}=\vec{r}_i-\vec{r}_j$, and $\hat{r}_{ij}=\vec{r}_{ij}/r_{ij}$. $F_{ij}^D$ and $F_{ij}^R$ are the dissipative and random force contributions respectively, where $\zeta$ is the drag coefficient, $\vec{v}_{ij}=\vec{v}_i-\vec{v}_j$, $k_B$ and $T$ are Boltzmann’s constant and the temperature respectively, and $\theta_{ij}$ is an uncorrelated Gaussian distributed random variable with zero mean and unit variance. $w_C$ and $\tilde{w}_C$ are weight functions, $\tilde{\rho}_i$ is a weighted density, and finally $r_c$ and $r_d$ are cutoff radii which set the range of the forces. The reason for introducing two cutoff radii is that the range of the density dependent repulsion must be smaller than that of the attraction (with $A_{ij} < 0$ and $B_{ij} > 0$) in order to obtain liquid vapor coexistence \citep{warren2003vapor}.

The polymer brush consists of endgrafted chains at varying grafting densities. The chains are grafted to a purely repulsive surface modeled using the Weeks-Chandler-Anderson (WCA) potential \citep{weeks1971role}. The lubricant is modeled as free chains. The free and grafted chains are all taken to be of the same species and therefore have the same interaction parameters among each other and with the liquid particles. In some cases the number of liquid particles in the droplet is varied to study the effects of droplet size.

The simulations are performed in the NVT ensemble, using periodic boundary conditions in all directions. The unit of energy is set by $k_BT=1$, the unit of length by the cutoff distance of the DPD attraction $r_c=1$, and the mass unit by $m=1$ for all species. In the following, all quantities are given in these units. Other fixed parameters include $r_d=0.8;~\zeta=4.5;~B_{ij}\equiv B=40;~dt=10^{-3}$. For the substrate WCA potential we choose $\sigma_{WCA}=1,\epsilon=1$. With this choice of parameters and the thickness of the brush, the substrate does not affect the wetting behavior. The spring constant $k=20$ and equilibrium extension $r_0=1$ are chosen for the bond potential. The resulting bond length is $a\approx1.09$. All simulations are performed in the absence of any gravitational forces. The rest of the parameters are varied to study different aspects of the system.

The simulations are conducted using the HOOMD-Blue simulation package \citep{anderson2020hoomd,phillips2011pseudo}. All snapshot visualizations are made with the OVITO visualization package \citep{stukowski2009visualization}.


\subsection{System Preparation}

\begin{figure}[!htb]
	\begin{centering}
		\includegraphics[width=8cm]{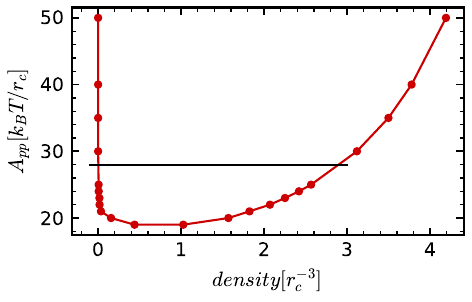}
	\end{centering}
	\caption{Melt-vapor coexistence phase diagram for a melt of lubricant chains of length $N_o=5$, the black line represents our choice of polymer-polymer cohesion $|A_{ij}|=|A_{pp}|=28$.}
	\label{fig:phaseDiag}
\end{figure}

The brush is composed of $n_B$ chains of length $N_B=50$ with the first monomer of each chain fixed on the substrate on a regular square lattice. We choose lattice constants $d=1.8-2.4$, corresponding to grafting densities $\sigma=0.17-0.31$. In addition, the system may contain $n_o$ lubricant polymers of length $N_o=5$. As part of an initial characterization of our system, we have determined the melt-vapor coexistence curve in a system containing lubricant only. Liquid-vapor phase separation sets in at $|A_{pp}|\approx 19$ (see Figure \ref{fig:phaseDiag}).

\begin{figure}[!htb]
	\begin{centering}
		\includegraphics[width=8cm]{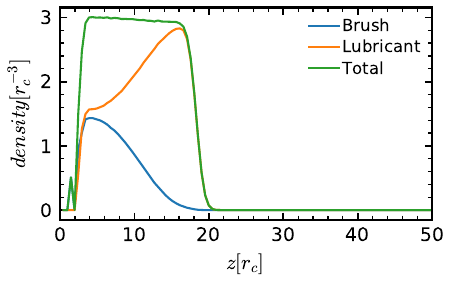}
	\end{centering}
	\caption{Example of density profiles for a polymer brush which is oversaturated with lubricant. The parameters are $|A_{pp}|=28, \sigma=0.25$, $n_B=900$ and $n_o=25\times10^3$}
	\label{fig:brushProfiles}
\end{figure}

The brush is first equilibrated without any lubricant for $8\times10^5$ simulation steps. We separately prepare a film of lubricant of length $N_o=5$ at the equilibrium melt density $\rho_o\approx2.9$ (see Figure \ref{fig:phaseDiag}). The lubricant is then added to the system by placing the film in contact with the brush and letting the lubricant infuse the brush. Equilibrium is reached after a maximum of $64\times10^5$ steps. Figure \ref{fig:brushProfiles} shows an example of density profiles in an equilibrated brush which is oversaturated with lubricant. The brush polymers exhibit the parabolic density profile which is characteristic for swollen brushes \citep{milner1988theory}. The overall density of the polymer film (brush and lubricant) is roughly constant throughout the whole film.

For the investigation of brush properties without a droplet we use brushes composed of $n_B = 30\times30$ chains. For brushes with spherical droplets we use $n_B = 100\times100$ chains \rev{resulting in typical box sizes of $200 \times 200 \times 300 \: r_c^3$ and lubricant content at the grafting density $\sigma = 0.25 \: r_c^{-2}$}. The number of liquid particles was $n_w =766\times10^3$ in most simulations, but was also sometimes reduced in order to study effects of the droplet size.

To prepare the spherical droplets we take a pendant droplet and place it in contact with the brush, and let it equilibrate in the absence of gravity. The systems are then left to equilibrate until no more conformational changes are observed. The systems with the droplet need at least $10^7$ steps to reach equilibrium.


\section{Results and Discussion}

\subsection{The Brush}

\label{sec:brush}

\begin{figure}
	\begin{centering}
		\includegraphics[width=8cm]{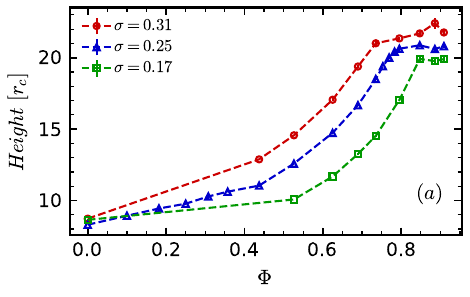}
        \includegraphics[width=8cm]{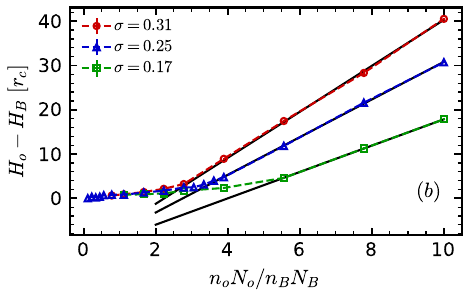}
	\par\end{centering}
	\caption{(a) Height of the brush versus the total fraction of lubricant $\Phi$ for different grafting densities and (b) difference in height between the brush and the top of the lubricant layer versus the ratio of lubricant particles to grafted chain particles $n_oN_o/n_BN_B$, for different grafting densities. The solid lines illustrate the extrapolation procedure used to determine the saturation point (see text). For $\sigma=0.25$, we obtain a saturation point $n_oN_o/n_BN_B\approx3.33$ or $\Phi\approx0.77$}
	\label{fig:HvNoli}
\end{figure}

To better understand the wetting behavior, we first investigate the characteristics of the brush. At grafting densities $\sigma\leq0.25$ the dry brush has a rough topography, with the density in the saturated sections having a value similar to the melt density. Figure \ref{fig:HvNoli} shows the variation in the height of the brush as a function of the amount of lubricant present for grafting densities $\sigma=0.17,0.25,0.31$. The x-axis is the fraction of lubricant in the system 
\begin{equation}
 \Phi=\frac{n_oN_o}{n_oN_o+n_BN_B},   
\end{equation}
where $n_o$ is the number of lubricant chains, $N_o$ is the length of lubricant chains, $n_B$ is the number of brush chains, and $N_B$ is the length of brush chains. This fraction does not always coincide with the fraction of lubricant strictly inside the brush, to which we will refer as $\Phi_o^B$, because a layer of lubricant forms on top of the brush if $\Phi$ exceeds a certain value (see below). \rev{With our choice of grafting density and length of the free chains, we are in the regime where the free chains are not expelled from the brush, and the film formed after the brush is saturated is stable against autophobic dewetting \citep{maas2002wetting,zhang2008wettability}.}\\

\begin{figure}[!htb]
	\begin{centering}
		\includegraphics[width=8cm]{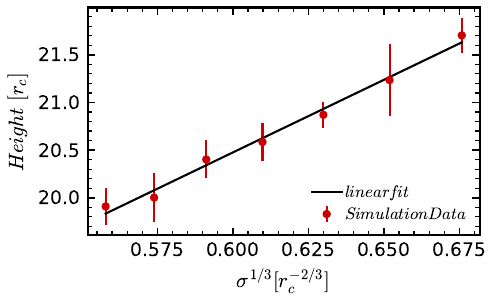}
	\end{centering}
	\caption{Height of the brush versus $\sigma^{1/3}$ for a saturated brush $\Phi=0.84$. The linear trend is consistent with the prediction for a brush in good solvent.}
	\label{fig:HvSig}
\end{figure}

We begin with examining the brush height, which we define as the point where the density distribution of brush particles drops below a cut-off value of $10^{-3}$. Here and in the rest of this article, the error bars correspond to the standard error on the average value obtained from 5 independent simulations. Figure \ref{fig:HvNoli}a shows that the brush height increases with the lubricant content as one might expect, but then reaches a constant value, indicating that the brush is saturated with lubricant. Once the brush is saturated, a film of lubricant forms on top of the brush, which is in coexistence with a vapor phase that has a very low lubricant density ($< 10^{-3}$). To identify the saturation point, we calculate the difference between the height of the brush ($H_B$) and the height of the lubricant ($H_o$), which are both defined as the height at which the respective densities drop below a value of $10^{-3}$. We then plot the difference, $H_o - H_B$, versus the ratio of the number of lubricant particles to brush particles $n_oN_o/n_BN_B$. When the brush is saturated it stops swelling, which means $H_o-H_B$ should increase linearly with the overall fraction of lubricant chains. Fitting the last 3 points of each curve for $H_o-H_B$ to a line, we can clearly detect the saturation point as the point where the curve deviates from that line. For example, at grafting density $\sigma=0.25$, the saturation point is at $\Phi\approx0.77$. This value will be chosen below to study droplets on brushes. Figure \ref{fig:HvSig} shows the variation of the brush height versus $\sigma^{1/3}$ for a saturated brush. The data points show linear behavior, which is consistent with a brush in good solvent \citep{de1980conformations}.
\\

For the cloaking and wetting properties, surface tension plays an important role. The surface tensions of the different components of the system are calculated through integration of the stress tensor anisotropy across flat interfaces 
\begin{equation}
  \gamma_x=\int dz(P_\perp-P_\parallel)  
\end{equation} 
Different configurations were simulated in order to determine separately the liquid-vapor interfacial tension $\gamma_{w}$, the lubricant-vapor tension $\gamma_{o}$, the lubricant-liquid tension $\gamma_{ow}$, the brush-vapor tension $\gamma_{b}$, and the brush-liquid tension $\gamma_{bw}$, A liquid or lubricant slab surrounded by vapor for $\gamma_{w}$ and $\gamma_{o}$, a brush for $\gamma_{b}$, a brush covered by a slab of liquid for $\gamma_{bw}$, and a lubricant slab in contact with a liquid slab for $\gamma_{ow}$. \rev{Here and throughout, the subscripts o, w, and b refer to the oil (lubricant), water (liquid droplet), and brush respectively. A single subscript refers to the interface of the respective phase with the vapor phase, while a combination of two subscripts refers to the interface between the two respective phases. A distinction needs to be made between the oil-water interface and the brush-water interface since the surface tension of the latter depends on the degree of swelling of the brush as will be shown below.}

Figure \ref{fig:PanisoVz} shows the stress anisotropy profile for an undersaturated and a fully saturated brush in contact with vapor, and an undersaturated brush in contact with liquid. The three curves show similar features, i.e., strong oscillations of  same magnitude near $z=0$ characterizing the brush-substrate interactions ($z<4$). This zone is followed by a zone where the stress anisotropy is slightly negative due to the mutual repulsion of the stretched polymers and decays to zero, and finally a positive peak at the position of the polymer interface away from the grafting surface. (In the case of the brush-liquid system, an additional peak appears around $z=40$, which marks the position of the liquid-vapor interface). 

Furthermore, we note that the curve for the strongly oversaturated brush covered by a lubricant layer (blue line corresponding to $\Phi=0.89$ in Figure \ref{fig:PanisoVz}) does not feature an additional peak which would mark the brush-lubricant interface. Hence 
we conclude that the interfacial tension between brush and lubricant film is zero. Furthermore, one can see that the behavior of the curves inside the brush is independent of the fraction of lubricant and of the type of the adjacent phase (vapor or liquid). Here we are mainly interested in the difference of $\gamma_{bw}$ and $\gamma_{b}$, which is the quantity entering the expected value for the equilibrium contact angle of the droplet (the Young's angle). Therefore, to reduce statistical errors, we choose to evaluate only the integral over the peaks at the interfaces, which correspond to the pure polymer-vapor and polymer-liquid interfacial tension, omitting the contributions of the brush and brush-substrate interactions to the surface energy. 

To be as close as possible to the experimental conditions of water on PDMS, we choose the interaction parameters such that the ratios of these surface tensions match the corresponding experimental values. Specifically, we choose $A_{ij}=A_{ww}=-50$ for the liquid-liquid cohesion which results in a liquid-vapor surface tension of $\gamma_{w}\approx3.2$, we choose $A_{ij}=A_{pp}=-28$ for the polymer-polymer cohesion which gives $\gamma_{o}\approx0.9$ for the pure lubricant-vapor interface, and finally $A_{ij}=A_{pw}=-21$ for the polymer-liquid cohesion which gives a lubricant-liquid surface tension of $\gamma_{ow}\approx1.4$.

\begin{figure}[!htb]
	\begin{centering}
		\includegraphics[width=8cm]{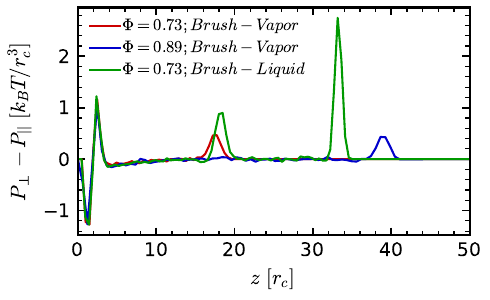}
	\end{centering}
	\caption{Stress tensor anisotropy for a brush with different lubricant content. The integral over the peak gives the relevant surface tension for the wetting. Both profiles show similar properties near the grafting surface, due to the repulsion of the surface and the tension in the first (grafted) bond. The large peak near \rev{$z=35$} is the liquid-vapor interface.}
	\label{fig:PanisoVz}
\end{figure}

\begin{figure}[!htb]
	\begin{centering}
		\includegraphics[width=8cm]{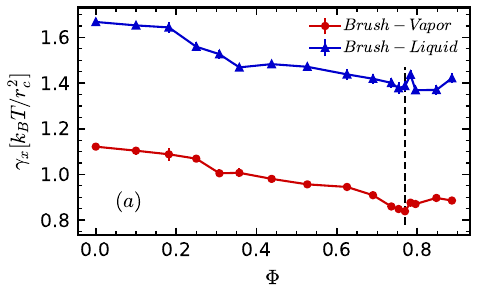}
        \includegraphics[width=8cm]{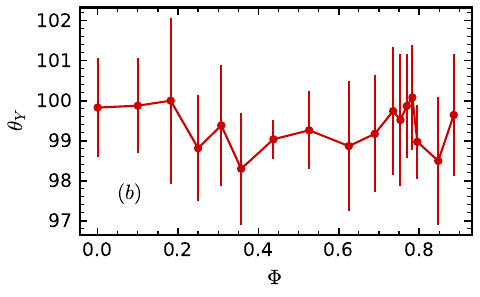}
	\end{centering}
	\caption{(a) Surface tension versus lubricant content for both the \rev{brush-vapor} and the \rev{brush-liquid} interfaces. (b) Young angle versus lubricant content, as calculated using equation \ref{eqn:young} with $\gamma_{w}=3.2$ }
	\label{fig:gamVNoli}
\end{figure}

The brush-vapor and brush-liquid surface tensions are calculated as a function of the lubricant content of the brush at $\sigma=0.25$. The results are shown in Figure \ref{fig:gamVNoli}a, while Figure \ref{fig:gamVNoli}b shows the predicted Young angles versus the lubricant content as calculated using equation (\ref{eqn:young}). The surface tension starts at a higher value for dry brushes, possibly due to the roughness, and then decreases and eventually reaches the melt (lubricant)  $\gamma_{o}$ value after saturation.



\subsection{Cloaking}
\label{sec:cloaking}

\begin{figure}
	\begin{centering}
		\includegraphics[width=8cm]{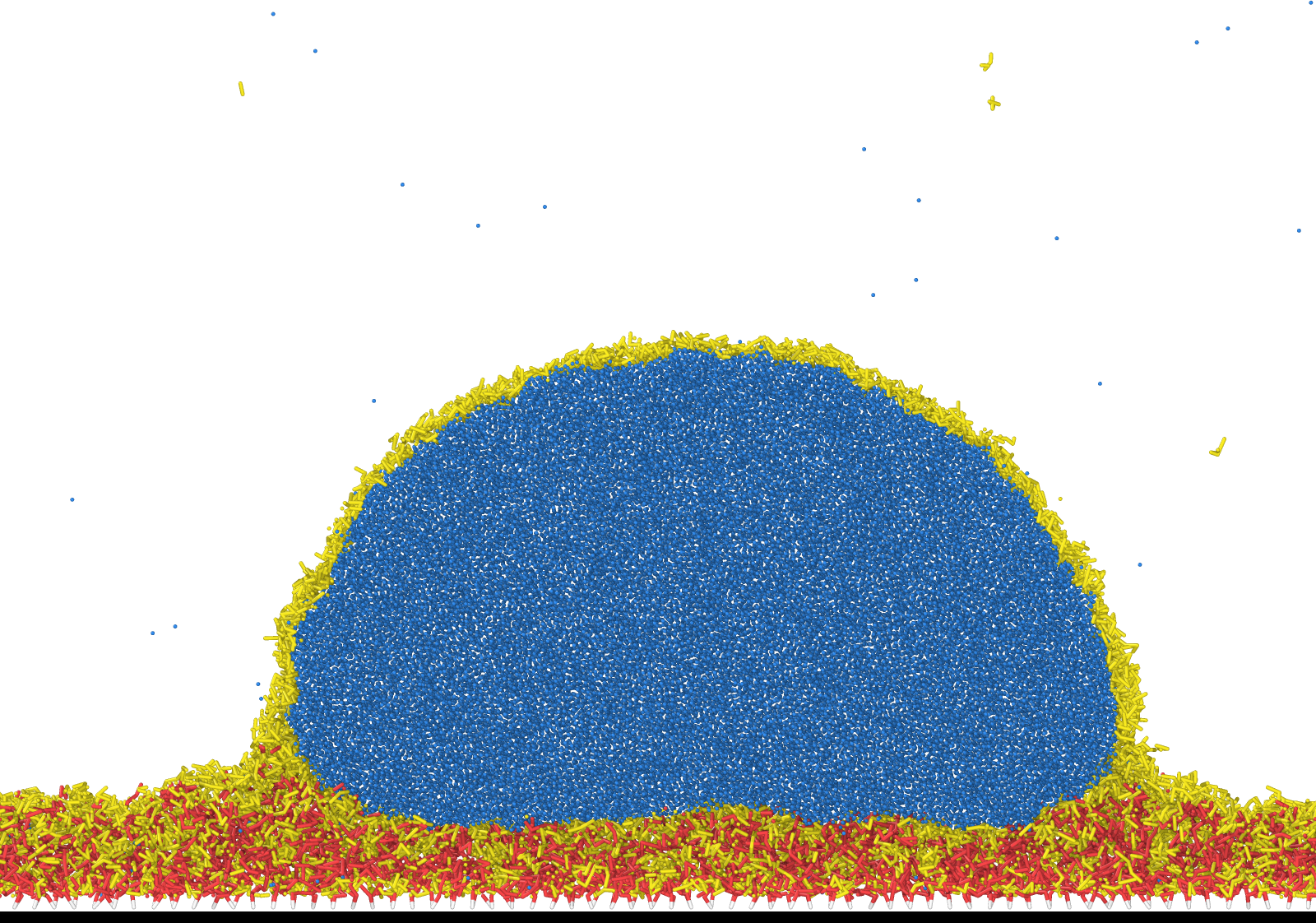}
	\end{centering}
	\caption{Side cross-sectional view of a spherical droplet on lubricated brush. Yellow particles represent the free chains, red the grafted chains, and blue the liquid particles.}
	\label{fig:sphDropLub} 
\end{figure}

\begin{figure}[!htb]
	\begin{centering}
		\includegraphics[width=8cm]{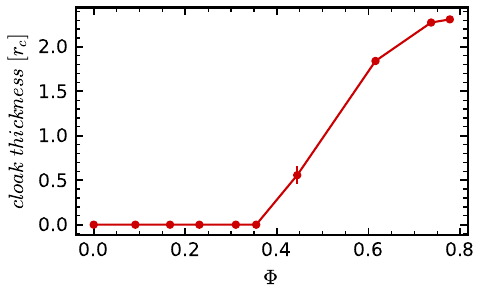}
	\end{centering}
	\caption{Cloak thickness as a function of lubricant content. $R$ in the legend is the radius of curvature of the droplet. We define a cloak as a layer of free chains that has the bulk density $\rho_o$ of the lubricant melt. The thickness is at zero before a certain transition point is reached, after which the thickness increases before seeming to flatten near the brush saturation point $\Phi\approx0.77$.}
	\label{fig:cloakVNoliCompare}
\end{figure}

When a water droplet is placed on a PDMS substrate, the residual silicone oil/lubricant accumulates on the droplet, resulting in a cloaking layer as shown in Figure \ref{fig:3DviewExp} and \citep{naga2021water}. The main driving force for this phenomenon is the positive spreading coefficient of silicone oil (uncrosslinked PDMS) on water $S_{ow}=\gamma_{w}-\gamma_{ow}-\gamma_{o}=11~\text{m}N/m$, effectively lowering the droplet-vapor surface tension. With our choice of parameters we get a spreading coefficient $S_{sim}=0.9$, which results in cloaking as shown in Figure \labelcref{fig:sphDropLub}. To investigate the cloaking, we study the cloak thickness as a function of the lubricant fraction in the system (see appendix \ref{app:cloaking} for details on the calculation method). We define a cloak as a layer of lubricant chains that is at the bulk density. Figure \ref{fig:cloakVNoliCompare} shows the cloak thickness versus lubricant content. The figure shows that cloaking is absent for low $\Phi$ and sets in at a certain transition point, beyond which the cloak thickness increases with the lubricant content. The graph seems to flatten as we approach the brush saturation point. No points beyond saturation were simulated here because of computational limitations.\\

\begin{figure}[!htb]
	\begin{centering}
		\includegraphics[width=8cm]{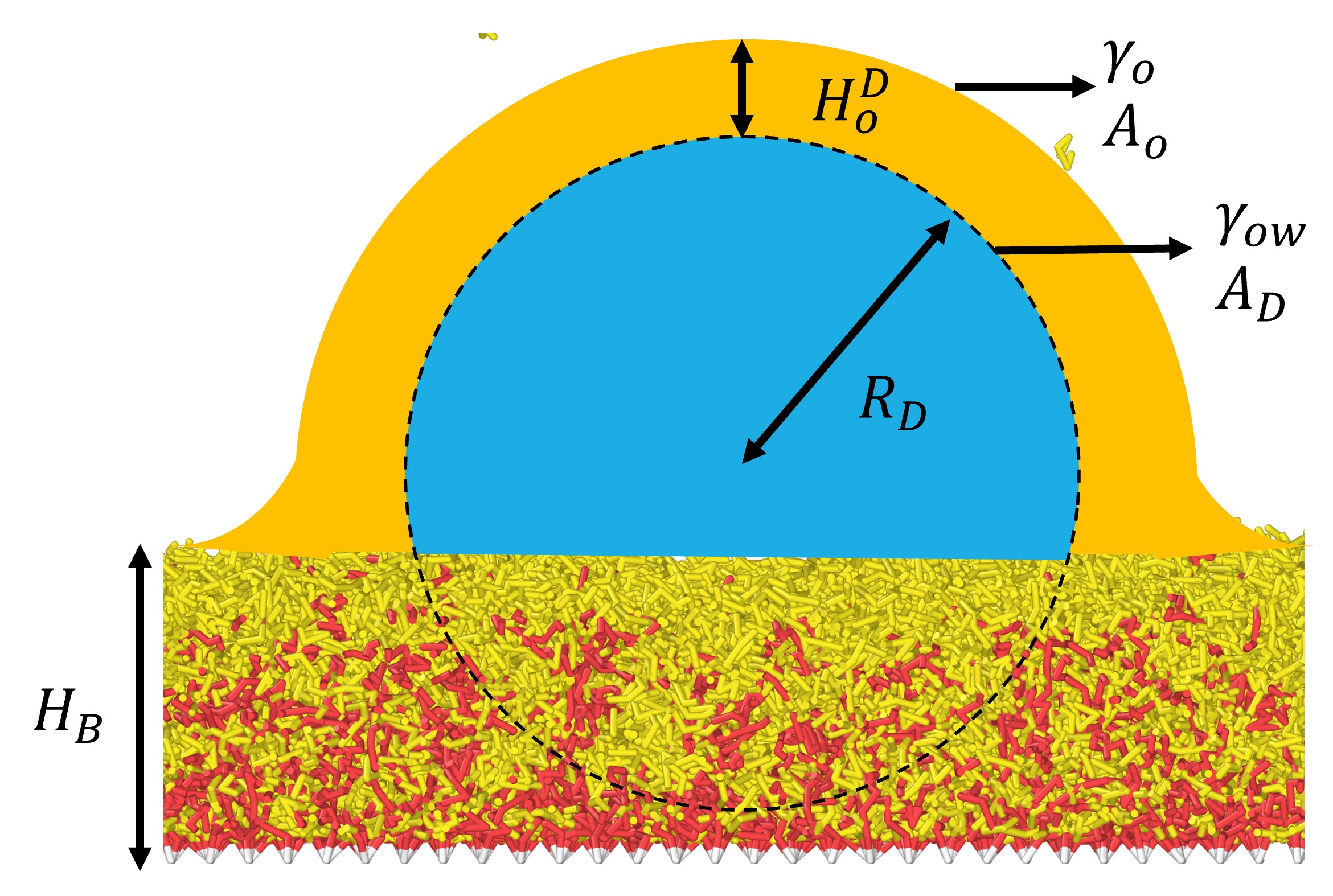}
	\end{centering}
	\caption{Diagram representing the most relevant quantities involved in the theory.}
	\label{fig:theoryDiag}
\end{figure}

To rationalize these findings, we develop a simple theoretical picture
(see appendix \ref{app:theory} for the full derivation\rev{, and Figure \ref{fig:theoryDiag} for a schematic representation}). We consider a coupled system of a spherical droplet, which is possibly covered by lubricant, and a lubricant infused brush, which can exchange lubricant chains. Equilibrium is achieved when the chemical potentials of both subsystems are equal. The free energy of the brush-lubricant subsystem is

\begin{equation}
    \label{eq:fb}
    \frac{F_B}{k_BT}=n_o\ln\Phi_o^B+ \frac{k}{2} \: n_B \: H_B^2-\mu_o^Bn_o
\end{equation}

\noindent where first term is the mixing free energy of the lubricant in the brush, the second term is the stretching elasticity cost of the brush with the spring constant $k/2 = K /N_B$ ($K$ is an undetermined prefactor with units of inverse length squared), and the last term is the chemical potential contribution. Eq. (\ref{eq:fb}) can be rewritten as the free energy per unit area and in terms of the fraction of lubricant in the brush as

\begin{multline}
     \tilde{F}_B=\frac{N_B}{N_o}\sigma\frac{\Phi_o^B\ln\Phi_o^B}{1-\Phi_o^B}+KN_B\sigma^3\frac{1}{(1-\Phi_o^B)^2}\\
     -\mu_o^B\frac{N_B}{N_o}\sigma \frac{\Phi_o^B}{1-\Phi_o^B}
\end{multline}

\noindent where $\Phi_o^B$ is the volume fraction of lubricant in the brush, and $\mu_o^B$ the chemical potential of the lubricant in the brush. If we impose the equilibrium condition $\partial\tilde{F}_B/\partial\Phi_o^B=0$, we can obtain a relation between the lubricant chemical potential and the lubricant content, $\Phi_o^B$, 

\begin{equation}
\label{eqn:muB}
    \mu_o^B=1-\Phi_o^B+\ln\Phi_o^B+2KN_o\sigma^2\frac{1}{1-\Phi_o^B}
\end{equation}

Next we consider the droplet-lubricant subsystem. 
In our simulations we work with short ranged forces. This leads us to hypothesize a rapidly decaying disjoining pressure for the cloaking layer, as opposed to the typical $h^{-3}$ dependence associated with long range Van der Waals forces. The driving mechanism for the cloaking phenomenon is the difference in surface tensions, captured by the spreading coefficient $S=\gamma_{w}-\gamma_{ow}-\gamma_{o}$. We then write our free energy contribution $F_\Pi$ of the disjoining pressure as

\begin{equation}
    F_\Pi=A_D \: S \: \exp{(-H_o^D/\xi)}
\end{equation}

\noindent where $H_o^D$ is the thickness of the cloaking layer, and $\xi$ a length scale characterizing the range of the disjoining pressure. Therefore, we write the free energy of the cloaking layer as

\begin{equation}
    \frac{F_D}{k_BT}=A_D S \exp{(-H_o^D/\xi)}+A_D\gamma_{ow}+A_o\gamma_{o}-\mu_o^D n_o^D
\end{equation}

\noindent The first term is the contribution of the disjoining pressure, the following two terms are the interfacial energies, and the last term is the contribution of the chemical potential. In the above expression  $\mu_o^d$ is the chemical potential of the lubricant on the droplet, $n_o^D$ the number of lubricant chains in the cloaking layer, $A_D$ and $A_o$ are the areas of the droplet surface and the surface of the cloaking layer, respectively, given by $$A_D=\kappa_D4\pi R_D^2$$ $$A_o=\kappa_o4\pi (R_D+H_o^D)^2,$$ and $\kappa_D$ and $\kappa_o$ are geometric factors that depend on the contact angle. For thin cloaking layers, we approximate $\kappa_D=\kappa_o\equiv\kappa$. Using this information we write the free energy of the cloaked droplet per unit area of the droplet $A_D$ as

\begin{equation}
    \tilde{F}_D= S \: \exp{(-H_o^D/\xi)}+\gamma_{ow}+\bigg(1+\frac{H_o^D}{R_D}\bigg)^2\gamma_{o}-\mu_o^D \frac{n_o^D}{\kappa4\pi R_D^2}
\end{equation}

Minimizing $\tilde{F}_D$ with respect to $n_o^D$, we obtain an expression for the relation between the chemical potential of the lubricant, $\mu_o^D$, and the thickness of the cloaking layer, $H_o^D$:

\begin{multline}
\label{eqn:muD}
    \mu_o^D=\frac{N_o}{\rho_o}\frac{1}{\bigg(1+\frac{H_o^D}{R_D}\bigg)^2}\bigg[\frac{-S}{\xi}\exp\bigg(-\frac{H_o^D}{\xi}\bigg)\\
    +2\frac{\gamma_o}{R_D}\bigg(1+\frac{H_o^D}{R_D}\bigg)\bigg].
\end{multline}

\noindent Equating equations (\ref{eqn:muB}) and (\ref{eqn:muD}) gives the equilibrium condition between the lubricated brush and the cloaked droplet $\mu_o^B(\Phi_o^B)=\mu_o^D(H_o^D)$. Setting $H_o^D=0$ on the R.H.S implies the existence of a transition when the fraction of lubricant is equal to the value $\Phi_o^{B*}$, which corresponds to a chemical potential value $\mu_o^{B*}$

\begin{equation}
    \mu_o^{B*}=\mu_o^B(\Phi_o^{B*})=\frac{N_o}{\rho_o}\bigg[-\frac{S}{\xi}+2\frac{\gamma_o}{R_D}\bigg]
\end{equation}

\noindent The existence of a cloaking transition is a clear feature from the results of our simulations. When the brush is fully saturated, including the configurations where a film of pure lubricant is formed on top of the brush, we have $\mu_o^B=0$. When this is the case the equilibrium condition is given by

\begin{equation}
\label{eqn:limH}
    \mu_o^D=0\implies\frac{-S}{\xi}\exp\bigg(-\frac{H_o^D}{\xi}\bigg)+2\frac{\gamma_o}{R_D}\bigg(1+\frac{H_o^D}{R_D}\bigg)=0.
\end{equation}

\noindent This means that the cloaking layer will reach a limiting thickness as the brush gets fully saturated, found by solving equation (\ref{eqn:limH}) for $H_o^D$. We were not able to simulate systems with an oversaturated brush and a droplet due to computational limitations, but the trend in Figure \ref{fig:cloakVNoliCompare} seems to be flattening near the saturation point $\Phi\approx0.77$.


\subsection{Cloaking and Wetting: effect on contact angles and the wetting ridge}
\label{sec:wetting}

q
After having investigated the cloaking transition, we now turn to study the influence of cloaking on the wetting properties of droplets, in particular the contact angle and the properties of the wetting ridge.

\subsubsection{Contact Angles on a Dry Brush}
\label{sec:dropOnDry}

\begin{figure}
	\begin{centering}
		\includegraphics[width=8cm]{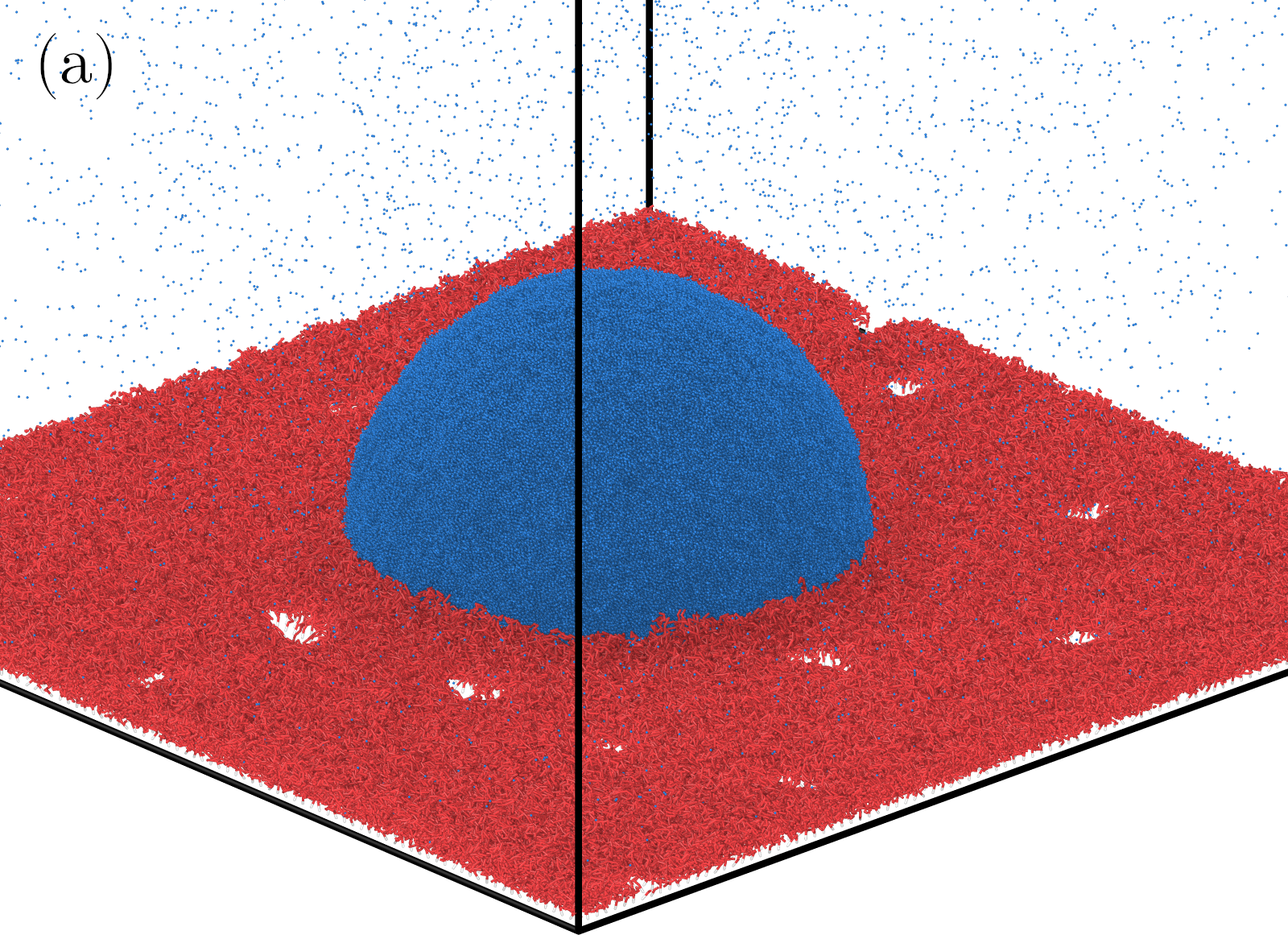}
        \includegraphics[width=8cm]{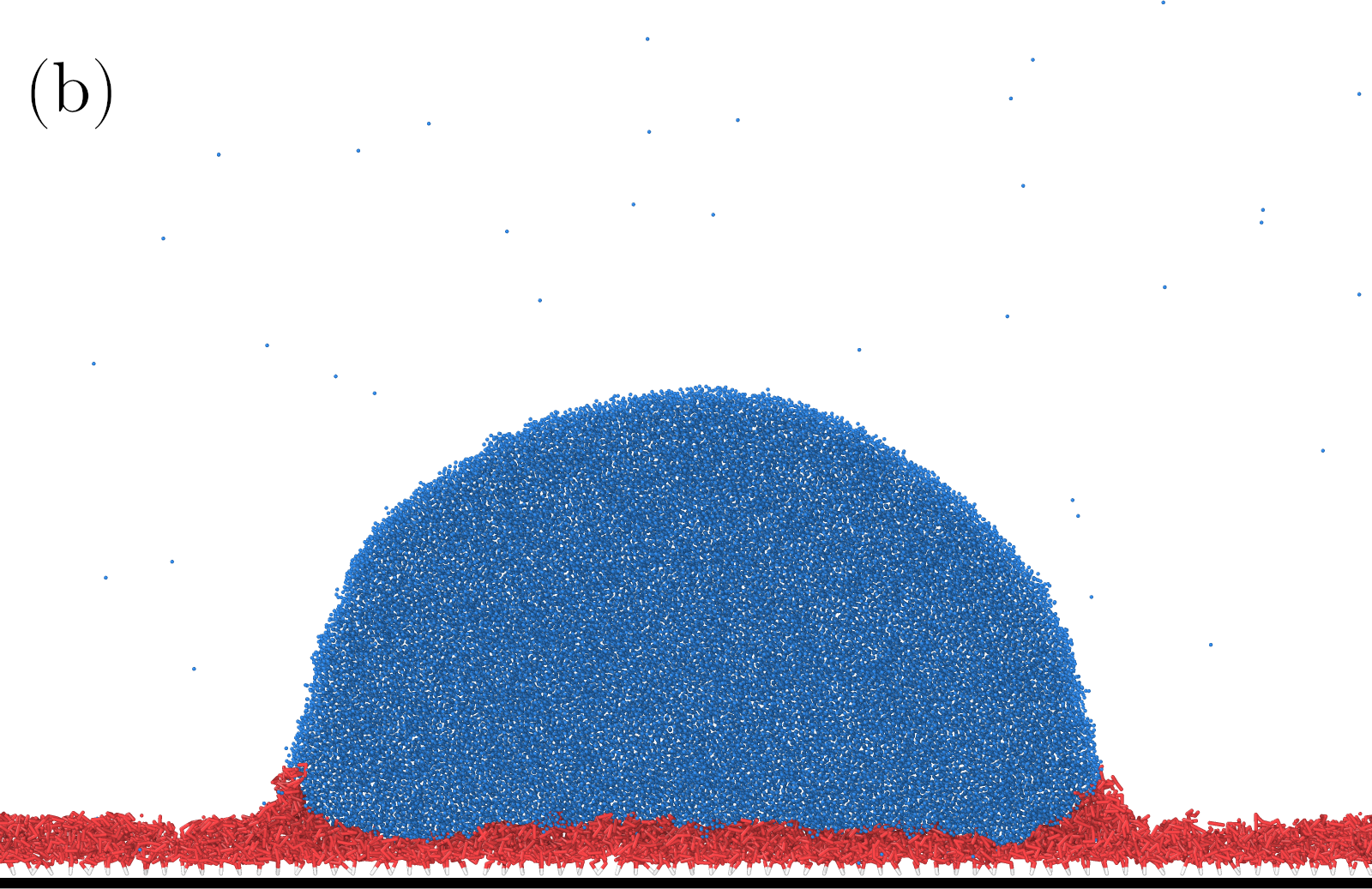}
		\par\end{centering}
	\caption{(a) Spherical Droplet on a dry brush (b) cross-sectional side view of the same droplet on a dry brush.}
	\label{fig:sphDrop} 
\end{figure}

We first investigate spherical droplets on dry brushes with a grafting density $\sigma=0.25$. This results in systems as shown in Figure \ref{fig:sphDrop}.

A key qualitative observation is the formation of a wetting ridge at the three phase contact line. Of particular interest is the contact angle between the droplet and the brush. Here we are specifically looking at the apparent contact angle, which can be compared to the angle calculated from the Young equation. The angle is calculated by fitting the top of the droplet to a circle and finding the angle it makes with a horizontal line at the level of the unperturbed brush. \rev{This procedure results in an apparent contact angle $\theta_{app}$.} For details on the calculation see Appendix \ref{app:contact}.\\

Figure \ref{fig:thVinvR} shows the variation of $\cos(\theta_{app})$ with the inverse radius of the contact line. The graph shows linear behavior, which is in agreement with the line tension hypothesis at the three phase contact line. The hypothesis suggests a contribution to the free energy that is proportional to the length of the three phase contact line. This results in a correction to the apparent contact angle of the form\cite{weijs2011origin}

\begin{equation}
	\cos(\theta_{app})=\cos(\theta_\infty)-\frac{\tau/\gamma_{w}}{r_{contact}}
\end{equation}

\noindent where $\theta_\infty$ is the limiting Young contact angle, achieved for infinitely large droplets, $\tau$ is the line tension, and $r_{contact}$ is the radius of the three phase contact line. Here the line tension is an effective parameter which possibly includes other corrections, most notably the correction due to the Tolman length. Fitting the data points to a line gives an effective line tension value $\tau/\gamma_{w}=-5.0$ and a Young angle $\theta_\infty = 101^\circ$.
The latter is in good agreement with the value calculated from the corresponding surface tensions\rev{, assuming the Shuttleworth effect is negligible due to the nature of the brush, where the stresses and strains are in the normal direction to the unperturbed brush interface\citep{shuttleworth1950surface}}.

\begin{figure}[!htb]
	\begin{centering}
		\includegraphics[width=8cm]{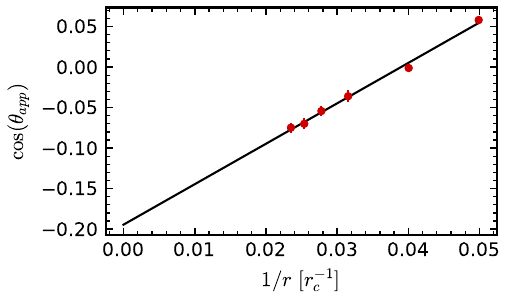}
	\end{centering}
	\caption{Variation of $\cos(\theta_{app})$ with the inverse radius of the contact line. The variation is linear suggesting that the dominant correction to the Young equation is that due to an effective line tension \rev{ $\cos(\theta_{app})=\cos(\theta_\infty)-\frac{\tau/\gamma_{w}}{r_d}$.} Our simulations suggest $\tau/\gamma_{w}=-5.0$ for a dry brush. The y-intercept has the value $-0.19$, which corresponds to a Young angle of $101^\circ$. This angle is in good agreement with the result calculated from surface tensions ($\approx100^\circ$)}
	\label{fig:thVinvR}
\end{figure}


\subsubsection{Contact Angles on a Lubricated Brush}
\label{sec:dropOnLub}

\begin{figure}[!htb]
	\begin{centering}
		\includegraphics[width=8cm]{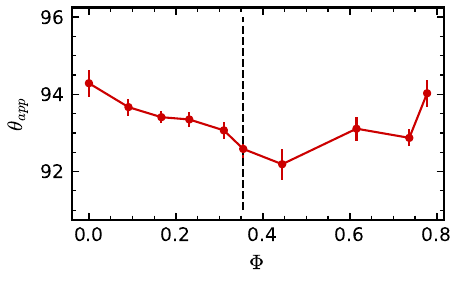}
	\end{centering}
	\caption{Variation of $\theta_{app}$ with the lubricant content. The contact angle decreases when lubricant is added and after the transition point (black dashed line), suggesting a possible increase in the magnitude of line tension.}
	\label{fig:thVNoliCompare}
\end{figure}

The next line of inquiry is to investigate the influence of lubricant on the contact angles. Figure \ref{fig:thVNoliCompare} shows the variation of the apparent contact angle on lubricated brushes. The graph shows a gradual decrease in the contact angle until about $\Phi=0.44$. This suggests the hypothesis that the magnitude of the line tension increases  when lubricant is added, and then again when the cloaking transition sets in, since in our case the line tension is negative and therefore promotes the spreading of the droplet.\\

\begin{figure}[!htb]
	\begin{centering}
		\includegraphics[width=8cm]{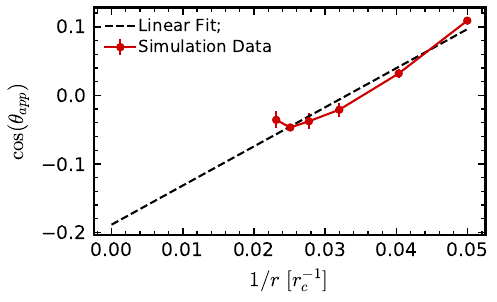}
	\end{centering}
	\caption{Variation of $\cos(\theta_{app})$ with the inverse radius of the contact line for a lubricated brush with $\Phi=0.44$. Here the behavior is not exactly linear, which is likely due to effects other than line tension. If we forcefully try to extract a line tension however, we get $\tau/\gamma_{w}=-5.7$ and the y-intercept has the value $-0.19$, which corresponds to a Young angle of $101^\circ$.}
	\label{fig:thVinvRLub}
\end{figure}

To test this hypothesis we measure the contact angle versus the size of the droplet on a lubricated brush beyond the cloaking transition point. The results are shown in Figure \ref{fig:thVinvRLub}. The points do not clearly fall on a line, indicating that the interpretation of the apparent contact angle in terms of surface tensions and line tensions may be too oversimplified, at least for the nanodroplets considered in this work.  If we nevertheless enforce a fit and try to extract the limiting angle and line tension, we get a value for the line tension $\tau/\gamma_{w}=-5.7$ and a limiting angle $\theta_\infty=101^\circ$. The line tension strength slightly increased, but more investigation is required to rule out other mechanisms that might be in play. 

After $\Phi=0.44$ we see a slight increase in the contact angle for the following two points. Since the droplet is cloaked for those points, the effective droplet-vapor surface tension is expected to be lower \citep{naga2021water}. Lowering the droplet surface tension drives the contact angle away from $90^\circ$, which is what we observe. The final point suggests a jump in apparent contact angle. However, this jump is most likely an artifact of the large curvature of the droplet near the contact line at high lubrication.

\subsubsection{Brush Ridge Reduction}

\begin{figure}[!htb]
	\begin{centering}
		\includegraphics[width=8cm]{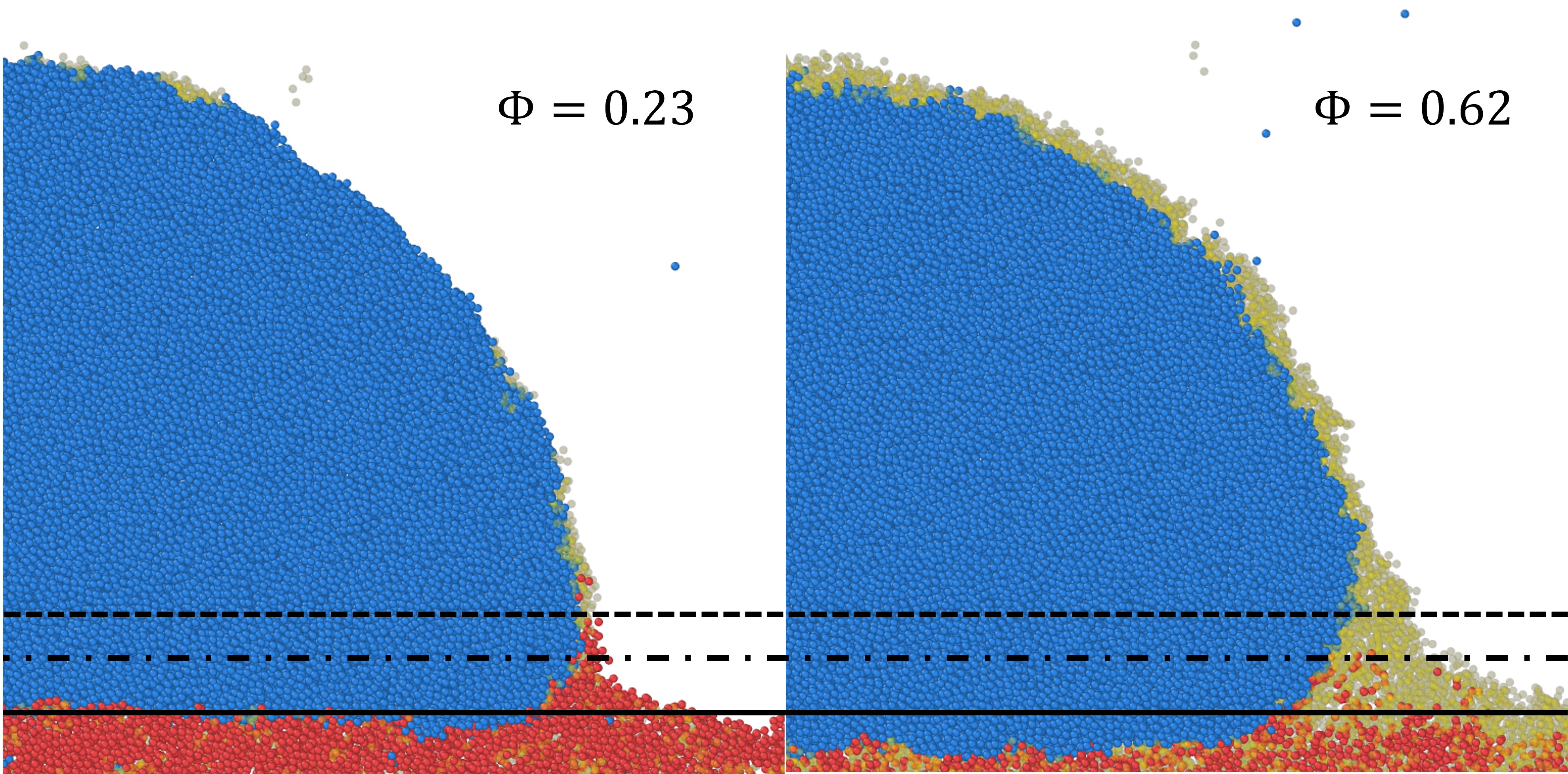}
	\end{centering}
	\caption{Snapshots showing the reduction in the grafted ridge height (red particles) as lubricant is added. The solid line shows the height of the unperturbed brush, the dashed line the height of the grafted ridge at low lubrication, and the dash-dot line the height at higher lubrication.}
	\label{fig:ridgeHeightClar}
\end{figure}

\begin{figure}[!htb]
	\begin{centering}
		\includegraphics[width=8cm]{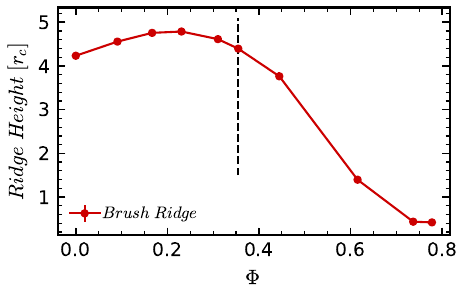}
	\end{centering}
	\caption{Height of the brush ridge versus the lubricant content, the dashed black line indicates the onset of the cloaking transition. The ridge height shows non-monotonic behavior. It first increases as lubricant is added, then decreases again, with the decrease being more steep after the onset of cloaking. }
	\label{fig:maxBHvNoliCompare}
\end{figure}

The cloaking of the droplet can be seen as an example for fluid separation from the underlying substrate, which is observed in certain cases of soft wetting. We observe in our simulation that the fluid separation is accompanied by a reduction in the height of the grafted chains near the droplet, in the vicinity of the wetting ridge as seen in Figure \ref{fig:ridgeHeightClar}. Such a phenomenon has also been observed experimentally for water droplets on swollen PDMS gels \cite{cai2021fluid}. Figure \ref{fig:maxBHvNoliCompare} shows the maximum height of the grafted chains near the droplet after subtracting the unperturbed height away from the droplet. The ridge height slightly increases after lubricant is added to the system. At some point it reaches a maximum then decreases. After cloaking sets in, the height decreases more rapidly as the the fraction of lubricant increases. The non-monotonic behavior before cloaking is qualitatively in agreement with the experimental results from \citep{cai2021fluid}.

\begin{figure}[!htb]
	\begin{centering}
		\includegraphics[width=8cm]{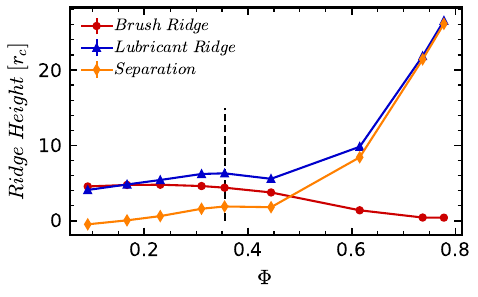}
	\end{centering}
	\caption{Variation of the lubricant ridge height, the brush ridge height, and the separation between the two with the lubricant content. In the absence of cloaking the ridge height is calculated directly. In the presence of cloaking it is defined as the height above the unperturbed brush where the cloak thickness increases by more than 10\% of its value in the region of constant thickness. The separation is calculated as the difference between the two ridge heights. }
	\label{fig:ridgeHeight}
\end{figure}

In many experiments, the resolution is not high enough to allow for visualization of the cloaking layer, but one can still observe a wetting ridge where Neumann angles can be calculated. The confocal image in Figure \ref{fig:3DviewExp} show an accumulation of lubricant close to the three phase contact line. This accumulation can be considered as an effective wetting ridge, so that if the resolution of the microscope were lower, the cloaking layer would not be visible, only the ridge. We observe a similar effect in our simulations, where the cloaking layer of constant thickness gets wider near the contact line (see Figure \ref{fig:rThDensOli}). From the simulation we can extract an effective ridge height, which we define as the position where the thickness of the cloaking layer increases by more than 10\% above its mean value far from the contact region, in the region of constant thickness. In the absence of a cloaking layer, the ridge height can be calculated directly. Figure \ref{fig:ridgeHeight} shows the change in the brush ridge height, the lubricant ridge height, and the lubricant-brush separation, as the amount of lubricant is increased. The lubricant-brush separation is calculated as the difference between the two ridge heights. The trends are very similar to the ones observed by Cai et al in \citep{cai2021fluid}. Notable is the rapid increase in the separation after the onset of cloaking.


\section{Conclusion}

In this study we have examined the cloaking of liquid droplets on lubricant infused polymer brushes by simulations and supporting experiments and theoretical considerations. 

The experiments clearly demonstrate the existence of such cloaking layers for droplets on brushes with a high lubricant content. This is accompanied by an accumulation of lubricant near the contact line between the droplet and the substrate. In experiments where the resolution does not allow for the observation of a cloaking layer, the accumulation near the contact line will appear as a Neumann type wetting ridge. The qualitative properties of the droplets in simulation are in very good agreement with the experiment\\

In the simulations we characterized the swelling of the brush by the lubricant and found the saturation point for different grafting densities, which plays a role in the thickness of the cloaking layer. The surface tension of the brush surface with vapor and liquid as a function of swelling is found to gradually decrease as one goes from a dry brush to a saturated brush, where for the latter the value approaches the values corresponding to the lubricant with the relevant interface.  \\

Furthermore, we characterized the cloaking of the droplet by the lubricant by calculating the variation of the cloak thickness versus the amount of lubricant present in the system. We find that a cloaking-transition takes place already for lubricant concentrations that are significantly lower than those where the brush saturates with lubricant. The cloak thickness seems to flatten as one approaches the saturation point of the brush. We derive a thermodynamic theory that predicts the cloaking transition, the increase of the cloak thickness, and the limiting thickness at saturation. Given the approximate nature of the theory however we do not have quantitative agreement. An additional effect of the presence of lubricant is liquid-substrate separation at the three phase contact line. This is accompanied by a decrease in the height of the substrate in the wetting ridge as more lubricant is added, which is an effect that has also been observed experimentally for droplets on PDMS gels.\\

We finally investigate the contact angles of the droplet on the brush. On a dry brush, we study the dependence of contact angles on the size of the droplet and find that line tension plays an important role for spherical droplets, where we find the value $\tau/\gamma_w=-5.0$. On lubricated brushes, we find that the angle is reduced, implying that the addition of lubricant increases the magnitude of line-tension (negative in our case). Overall, the effect of cloaking on the apparent contact angle is remarkably small, in the range of only a few degrees. It is strongly affected by the finite size of the droplet, but the extrapolated value for infinite droplet size is in good agreement with the Young's angle predicted from the surface tension, both for cloaked and uncloaked droplets. Thus we find that the cloaking transition has only a minor effect on the static equilibrium wetting properties of droplets.\\

Given that cloaking has a strong effect on the structure of the contact line, we do however expect that it will strongly influence the dynamic wetting properties, i.e., contact angle hysteresis and the friction of rolling droplets. This will be the subject of future investigations. \rev{Another interesting subject for future investigation is the spreading and imbibition of the free chains into the brush and comparison with atomistic simulations \citep{etha2021wetting}.}


\section{Acknowledgments}

This work was funded by the German Science Foundation within the priority program SPP 2171 (grants Schm 985/22 and VO 639/16). R.B. thanks Leonid Klushin for useful discussions.

\newpage


\appendix

\section{Experimental Details}
\label{app:exp}


\subsection{PDMS brush substrate}
Glass slides ($24 \times 60~mm$, $170~\mu m$ thick) were rinsed with ethanol and acetone and dried under nitrogen gas. Thereafter, they were exposed to oxygen plasma (0.3 bar) for 10 minutes at $300~W$ (Femto, Diener Electronic, Germany). This treatment adds hydroxyl groups to the surface. Approximately $250~\mu l$ of methyl terminated PDMS oil ($6000~Da$) is cast on the surface and distributes homogeneously. Over time, PDMS chains condense on the hydroxyl (anchor) groups at the surface \cite{teisala2020grafting}. After 24h, we remove uncrafted PDMS chains from the surface by immersing the substrate in a toluene-sonication bath for 15 minutes. A $5-10~nm$ thick film of surface-grafted PDMS chains remains. Free PDMS oligomers ($3500~Da$) are distributed overnight ($3 \pm 2~\mu l/mm^2$) into the grafted PDMS chains. The oligomers were fluorescently labeled (Lumogen Red, BASF, $0.1~mg/ml$).


\subsection{Experimental visualization of oligomer cloaking}
We place the PDMS brush substrate on a laser scanning confocal microscope (Leica TCS SP8)). We place a droplet (approx. $500~nl$) of glycerol ($57~\%wt$) water ($43~\%wt$) on the substrate. The glycerol-water mixture was chosen to suppress evaporation. The mixing ratio yields a refractive index match between drop and PDMS (i.e., $n=1.4$) and a surface tension of around $\gamma_{w}\approx67~mN/m$\cite{brindise2018density,takamura2012physical}. Combining this with $\gamma_{o}\approx21~\text{m}N/m$ for silicone-oil-air, and $\gamma_{wl}\approx40~\text{m}N/m$ for silicone-oil-water, we expect an apparent contact angle (for uncloaked droplets) $\theta_Y=\cos^{-1}\bigg(\frac{\gamma_{o}-\gamma_{wl}}{\gamma_{w}}\bigg)\approx106^\circ$.

After 20 minutes, we capture 3D stacks of the droplet using a Leica HC
PL APO CS 10x/0.40 objective lens. We observe stark silicone oil (yellow) accumulations in the vicinity of the three-phase contact line in Figure \ref{fig:3DviewExp}. Additionally, silicone oil distributes on the entire air-exposed interface of the drop, indicated by the fine fluorescence signal. This gives direct indication of a silicone oil cloak on the droplet.

\section{Calculation Methods}


\subsection{Contact Angle Calculation}
\label{app:contact}

\begin{figure}[!htb]
	\begin{centering}
		\includegraphics[width=8cm]{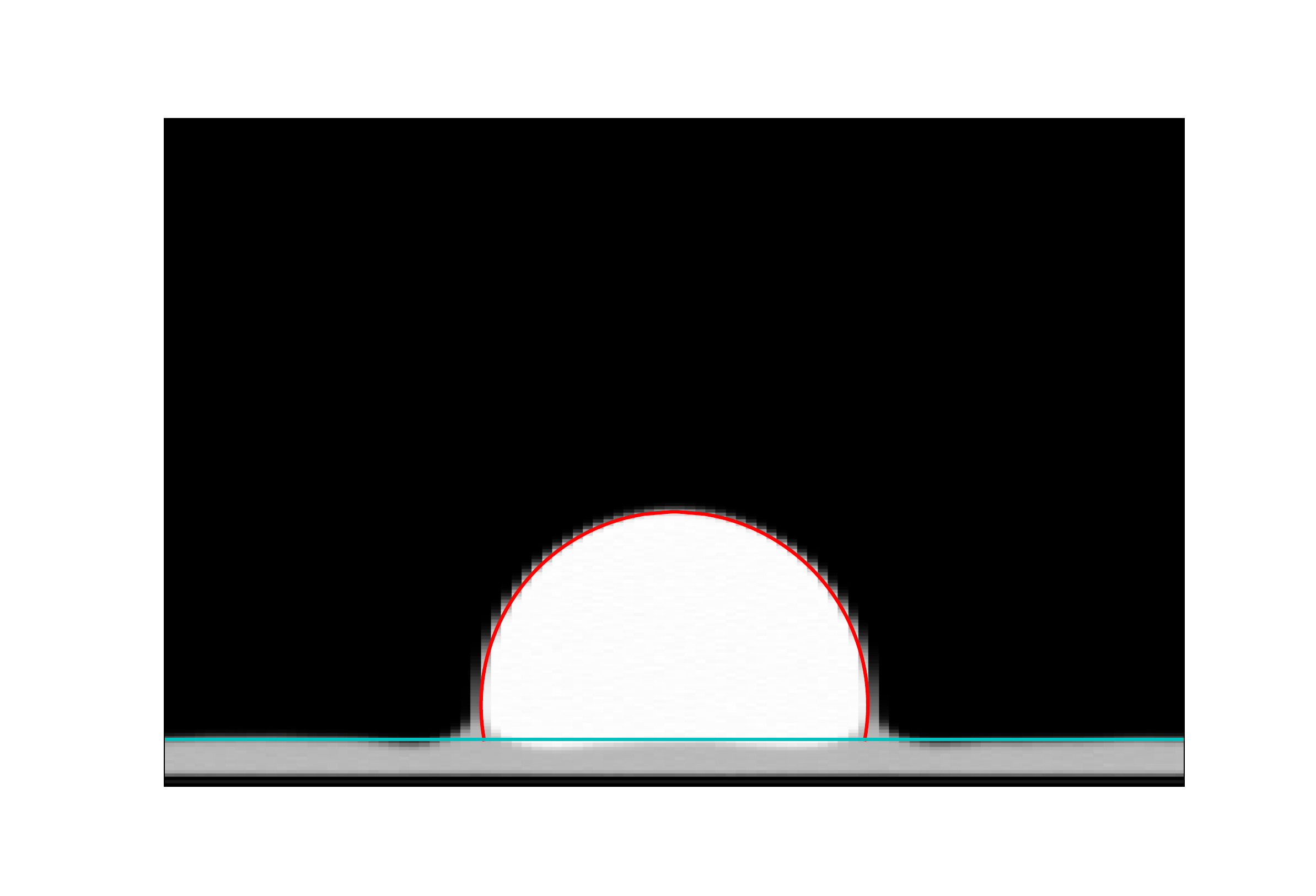}
	\end{centering}
	\caption{Density map of a droplet on a brush, which also show the circular fit of the droplet (red) and the level with which the apparent angle is measured (cyan)}
	\label{fig:dryBrushFit}
\end{figure}

 The density of the different components of the systems is calculated as function of position by dividing the simulation box into bins. The density is averaged over the azimuthal direction. From the density map, the interface of a specific comonent is defined as the level curve where the density of that component reaches half its bulk value. The contact angle is measure by fitting the droplet-vapor interface to a circle and calculating the angle this circle makes with the surface of the brush, as shown in Figure \ref{fig:dryBrushFit}. It is important to clarify here that the surface of the brush is the surface of grafted chains, to have consistent calculations when the brush is over-saturated.
 
\subsection{Cloak Thickness Calculation}
\label{app:cloaking}

In order to find the cloak thickness the density of the lubricant above the brush height was calculated as a function of the radial position and angle from the z-axis. The origin is chosen at the center of curvature for the droplet as calculated in Appendix \ref{app:contact}. This results in a density map $\rho(r,\theta)$. We define a cloak as a layer at the bulk density $\rho_o$ of lubricant and so the level curves of $\rho=\rho_o/2$ were found and plotted. One can clearly see that the thickness of the cloak is the same everywhere on the droplet except when one approaches the substrate. The cloak thickness was then calculated as the average distance between the lines for $\theta\in[0,\pi/6]$. This range was chosen since the thickness was observed to be constant for all chosen lubrication values.

\begin{figure}[!htb]
	\begin{centering}
		\includegraphics[width=8cm]{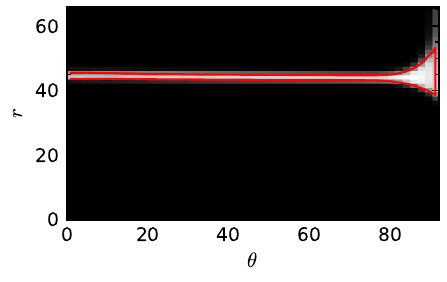}
	\end{centering}
	\caption{Density map of lubricant versus the radial position from the center of curvature of the droplet (y-axis) and the angle from the z-direction (x-axis). The red line represents the level curve of $\rho=\rho_o/2$. The cloak thickness is calculated as the distance between the red lines where they are parallel.}
	\label{fig:rThDensOli}
\end{figure}


\section{Derivation of the thermodynamic theory for cloaking}
\label{app:theory}


\subsection{Notation}
\begin{itemize}
    \item $k_B$: Boltzmann constant
    \item $T$: temperature
    \item $n_o^B$: number of lubricant chains
    \item $N_o$: length of a lubricant chain
    \item $n_B$: number of grafted chains
    \item $N_B$: length of a grafted chain
    \item $a$: size of a monomer
    \item $\Phi_o^B$: fraction of lubricant in the brush
    \item $H_B$: height of the brush
    \item $\mu_o^B$: chemical potential of the lubricant in the brush
    \item $\sigma$: grafting density of the brush
    \item $\gamma_{w}$: surface tension of the droplet-vapor interface
    \item $\gamma_{o}$: surface tension of the lubricant-vapor interface
    \item $\gamma_{ow}$: surface tension of the droplet-lubricant interface
    \item $S=\gamma_{w}-\gamma_{ow}-\gamma_{o}$: spreading coefficient of the lubricant on the lubricant on the droplet
    \item $H_o^D$: thickness of the lubricant layer
    \item $\xi$: length scale of the decay of the disjoining pressure
    \item $R_D$: radius of curvature of the droplet
    \item $\mu_o^D$: chemical potential of the cloaking layer (lubricant on the droplet)
    \item $n_o^D$: number of lubricant chains on the droplet
    \item $\rho_o$: density of pure lubricant melt
\end{itemize}


\subsection{Free energy of the brush}

We consider the lubricated brush and the cloaked droplet as coupled thermodynamic systems which can exchange energy and lubricant chains. The free energy of the brush subsystem is given by

\begin{equation}
    \frac{F_B}{k_BT}=n_o\ln\Phi_o^B+Kn_B\frac{H_B^2}{N_B}-\mu_o^Bn_o
\end{equation}

\noindent The first term is the mixing free energy of the lubricant in the brush, the second term is the stretching elasticity cost of the brush the spring constant $k/2 = K /N_B$ ($K$ is an undetermined prefactor with units of inverse length squared), and the last term is the chemical potential contribution. Here $\mu_o^B\leq0$ with $\mu_o^B=0$ corresponding to a fully saturated brush. Rewriting $n_o^B$ and $n_B$ as $$n_o^B=A_BH_B\frac{\Phi_o^B}{N_oa^3}$$ $$n_B=A_B\sigma$$ with $A_B$ the surface area of the brush, we can rewrite the free energy as

\begin{equation}
    \frac{F_B}{k_BT}=A_BH_B\frac{\Phi_o^B}{N_oa^3}\ln\Phi_o^B+A_B\sigma K\frac{H_B^2}{N_B}-\mu_o^BA_BH_B\frac{\Phi_o^B}{N_oa^3}
\end{equation}

\noindent The fraction of grafted chains is given by

\begin{equation}
    \Phi_B=1-\Phi_o^B=\frac{N_B\sigma a^3}{H_B}
\end{equation}

\noindent which leads to an expression for the height of the brush

\begin{equation}
    H_B=\frac{N_B\sigma a^3}{1-\Phi_o^B}
\end{equation}

\noindent making this final replacement and dividing by the area $A_B$ to get the free energy per area we get the final form for our free energy

\begin{multline}
     \tilde{F}_B\equiv\frac{F_B}{k_BTA_B}=\frac{N_B}{N_o}\sigma\frac{\Phi_o^B\ln\Phi_o^B}{1-\Phi_o^B}+KN_B\sigma^3\frac{1}{(1-\Phi_o^B)^2}\\
     -\mu_o^B\frac{N_B}{N_o}\sigma \frac{\Phi_o^B}{1-\Phi_o^B}
\end{multline}

\noindent Where a factor of $a^6$ has been absorbed into the prefactor $K$. Now if we impose the equilibrium condition $\partial\tilde{F}_B/\partial\Phi_o^B=0$, we can obtain the chemical potential as

\begin{equation}
    \mu_o^B=1-\Phi_o^B+\ln\Phi_o^B+2KN_o\sigma^2\frac{1}{1-\Phi_o^B}
\end{equation}


\subsection{Free energy of the cloak}

In our simulations we work with short ranged forces. This leads us to hypothesize a rapidely decaying disjoining pressure for the cloaking layer, as opposed to the typical $h^{-3}$ dependence associated with long range Van der Waals forces. Given that the driving mechanism for the cloaking phenomenon is the difference in surface tensions, captured by the spreading coefficient $S=\gamma_{w}-\gamma_{ow}-\gamma_{o}$, we write our free energy contribution $F_\Pi$ of the disjoining pressure as

\begin{equation}
    F_\Pi=A_D\: S\: \exp{(-H_o^D/\xi)}.
\end{equation}

\noindent Therefore, we write the free energy of the cloaking layer as

\begin{equation}
    \frac{F_D}{k_BT}=A_D\: S\: \exp{(-H_o^D/\xi)}+A_D\gamma_{ow}+A_o\gamma_{o}-\mu_o^D n_o^D
\end{equation}

\noindent and dividing by the area of the droplet we get

\begin{equation}
    \tilde{F}_D=\frac{F_D}{k_BTA_D}=S \: \exp{(-H_o^D/\xi)}+\gamma_{ow}+\frac{A_o}{A_D}\gamma_{o}-\mu_o^D \frac{n_o^D}{A_D}
\end{equation}

\noindent where $A_D$ and $A_o$ are the areas of the droplet and the cloaking layer respectively, given by $$A_D=\kappa_D4\pi R_D^2$$ $$A_o=\kappa_o4\pi (R_D+H_o^D)^2$$ where $\kappa_D$ and $\kappa_o$ are geometric factors that depend on the contact angle. For thin cloaking layers, we approximate $\kappa_D=\kappa_o\equiv\kappa$. Inserting this into the expression for $\tilde{F}_D$ we get

\begin{equation}
    \tilde{F}_D=\: S\:\exp{(-H_o^D/\xi)}+\gamma_{ow}+\bigg(1+\frac{H_o^D}{R_D}\bigg)^2\gamma_{o}-\mu_o^D \frac{n_o^D}{\kappa4\pi R_D^2}
\end{equation}

\noindent To minimize $\tilde{F}_D$ with respect to $n_o^D$, we use

\begin{equation}
    n_o^D=\frac{\rho_o}{N_o}V_o\implies dn_o^D=\frac{\rho_o}{N_o}dV_o
\end{equation}

\noindent where $V_o$ is the volume of the cloaking layer, whose differential $dV_o$ is given by

\begin{equation}
    dV_o=\kappa4\pi(R_D+H_o^D)^2 dH_o^D
\end{equation}

\noindent the above two equations give

\begin{equation}
    \frac{\partial H_o^D}{\partial n_o^D}=\frac{N_o}{\rho_o}\bigg[\kappa4\pi(R_D+H_o^D)^2\bigg]^{-1}
\end{equation}

\noindent using the above equation to evaluate $\partial\tilde{F}_D/\partial n_o^D=0$ we obtain the chemical potential of the cloaking layer as

\begin{equation}
    \mu_o^D=\frac{N_o}{\rho_o}\frac{1}{\bigg(1+\frac{H_o^D}{R_D}\bigg)^2}\bigg[\frac{-\: S\:}{\xi}\exp\bigg(-\frac{H_o^D}{\xi}\bigg)+2\frac{\gamma_o}{R_D}\bigg(1+\frac{H_o^D}{R_D}\bigg)\bigg]
\end{equation}


\subsection{Cloaking transition and limiting cloak thickness}

Equilibrium between the brush and the cloaked droplet is achieved when the chemical potentials are equal $\mu_o^B(\Phi_o^B)=\mu_o^D(H_o^D)$. The transition from uncloaked to cloaked droplet is found by setting $H_o^D=0$ on the R.H.S implying the existence of a transition when the fraction of lubricant is equal to the value $\Phi_o^{B*}$ which corresponds to a chemical potential value $\mu_o^{B*}$

\begin{equation}
    \mu_o^{B*}=\mu_o^B(\Phi_o^{B*})=\frac{N_o}{\rho_o}\bigg[-\frac{\: S\:}{\xi}+2\frac{\gamma_o}{R_D}\bigg]
\end{equation}

\noindent when the brush is fully saturated, including the configurations where a film of pure lubricant is formed on top of the brush, we have $\mu_o^B=0$. When this is the case the equilibrium condition is given by

\begin{equation}
    \mu_o^D=0\implies\frac{-\: S\:}{\xi}\exp\bigg(-\frac{H_o^D}{\xi}\bigg)+2\frac{\gamma_o}{R_D}\bigg(1+\frac{H_o^D}{R_D}\bigg)=0
\end{equation}

\noindent which means that the cloaking layer will reach a limiting thickness as the brush gets fully saturated.


%
%

\newpage

\bibliographystyle{unsrt}
\bibliography{refs}

\end{document}